\documentclass[11pt,preprint]{aastex}

\begin{document}

\shorttitle{Truncated Disks around HD 98800 and Hen 3-600}

\shortauthors{Andrews et al.}

\title{Truncated Disks in TW Hya Association Multiple Star Systems}

\author{Sean M. Andrews\altaffilmark{1,2}, Ian Czekala\altaffilmark{1,3}, D. J. Wilner\altaffilmark{1}, \\ Catherine Espaillat\altaffilmark{1,4}, C. P. Dullemond\altaffilmark{5}, and A. M. Hughes\altaffilmark{1}}

\altaffiltext{1}{Harvard-Smithsonian Center for Astrophysics, 60 Garden St., Cambridge, MA 02138; sandrews@cfa.harvard.edu}
\altaffiltext{2}{Hubble Fellow}
\altaffiltext{3}{University of Virginia Department of Astronomy, P. O. Box 400325, Charlottesville, VA 22904}
\altaffiltext{4}{NSF Astronomy \& Astrophysics Postdoctoral Fellow}
\altaffiltext{5}{Max Planck Institut f\"{u}r Astronomie, K\"{o}nigstuhl 17, 69117 Heidelberg, Germany}

\begin{abstract}
We present high angular resolution (down to $0\farcs3 \approx 13$\,AU in 
diameter) Submillimeter Array observations of the 880\,$\mu$m (340\,GHz) 
thermal continuum emission from circumstellar dust disks in the nearby 
HD\,98800 and Hen\,3-600 multiple star systems.  In both cases, the dust 
emission is resolved and localized around one stellar component $-$ the 
HD\,98800\,B and Hen\,3-600\,A spectroscopic binaries $-$ with no evidence for 
circum-system material.  Using two-dimensional Monte Carlo radiative transfer 
calculations, we compare the SMA visibilities and broadband spectral energy 
distributions with truncated disk models to empirically locate the inner and 
outer edges of both disks.  The HD\,98800\,B disk appears to be aligned with 
the spectroscopic binary orbit, is internally truncated at a radius of 3.5\,AU, 
and extends to only 10-15\,AU from the central stars.  The Hen\,3-600\,A disk 
is slightly larger, with an inner edge at $\sim$1\,AU and an outer radius of 
15-25\,AU.  These inferred disk structures compare favorably with theoretical 
predictions of their truncation due to tidal interactions with the stellar 
companions.
\end{abstract}

\section{Introduction}

A large fraction of the stars in the solar neighborhood are members of 
gravitationally bound multiple systems \citep{abt76,duquennoy91,lada06}.  
Similar or higher multiplicity fractions are found in nearby young stellar 
associations \citep{ghez93,leinert93,simon95}, confirming that multiplicity 
is a common outcome in any star formation environment.  With such a substantial 
number of stars residing in multiple systems, it is natural to also consider 
the potential impacts of stellar companions on the planet formation process.  
The structures of the circumstellar disks used to build planets can be strongly 
affected by tidal interactions with the stellar components in these systems 
\citep{lin93,artymowicz94,pierens08}.  These interactions can open gaps in 
disks, truncate their edges, and potentially expedite their dissipation before 
planetesimals can be made.  But remarkably, despite these formidable dynamical 
hazards, a relatively large fraction ($\sim$25\%) of exoplanet host stars have 
stellar companions \citep[][and references therein]{raghavan06,desidera07}.  
Observations of circumstellar material in young multiple star systems can be 
used to empirically characterize the impact of these tidal interactions on disk 
structures, and therefore to place constraints on when and where planet 
formation can potentially be accommodated in the presence of a stellar 
companion.  

The fate of the circumstellar material in a young multiple star system is 
primarily dependent on the orbital separation ($a$), eccentricity ($e$), and 
mass ratio ($\mu$) of the stellar components \citep{artymowicz94}.  Systems 
with large separations ($a \sim$ hundreds of AU) impart little or no dynamical 
effects on their circumstellar material, leaving disks around each stellar 
component that are similar to those around single stars.  Conversely, 
individual disks in a small-separation system ($a \sim$ a few AU or less) will 
likely not survive.  Instead, these systems can host a single circum-system 
disk with a dynamically cleared central cavity out to a radius comparable to 
the stellar separation ($\sim$2-3$a$).  Qualitatively, such theoretical 
predictions have been broadly supported by observations.  The disks around 
small-separation spectroscopic binaries are like those around single stars 
except for a diminished dust emission signal shortward of $\sim$5\,$\mu$m, 
consistent with a lack of warm material inside a few AU \citep{jensen97}.  
Likewise, disks in wide binary systems are comparable to their more isolated 
counterparts, although the secondary disk signatures are often weak or absent 
\citep[e.g.,][]{jensen03,aw05b,patience08}.  

But most multiple systems in both the field and young clusters have orbits with 
intermediate separations \citep[$a \sim$ tens of AU; e.g.,][]{mathieu00}.  The 
disks in these systems suffer the most dramatic effects of star-disk 
interactions, resulting in their external truncation at a fraction of the 
component separation ($\sim$0.2-0.5$a$), or perhaps their complete dispersal.  
Single-dish radio photometry surveys have demonstrated that the $\sim$1\,mm 
luminosities for medium-separation binaries are significantly lower than their 
more closely or widely separated counterparts \citep[as well as single 
stars;][]{jensen94,jensen96,osterloh95,andrews05}.  Since the dust emission at 
these wavelengths is optically thin \citep{beckwith90}, this luminosity deficit 
is interpreted as indirect evidence for a diminished disk mass due to the tidal 
truncation of the outer disk.  However, this {\it total mass} deficit does not 
necessarily have any bearing on the potential for making planets around 
medium-separation multiples.  The likelihood for planet formation depends more 
on reaching a critical disk {\it density}, and therefore it is more important 
to identify how that total mass is distributed spatially in such systems.  
Doing so involves locating where the disks in multiple star systems are 
truncated using high angular resolution observations of their millimeter 
continuum emission 
\citep[e.g.,][]{jensen96b,akeson98,guilloteau99,guilloteau08}.

In this article, we present the first set of high angular resolution millimeter 
observations of the circumstellar material in two nearby multiple star systems, 
HD\,98800 and Hen\,3-600.  Both systems are members of the TW Hya association 
and are therefore among the nearest pre-main sequence stars, with a distance of 
roughly 50\,pc \citep{kastner97,webb99}.  HD\,98800 is a rare quadruple star 
system, consisting of a pair of spectroscopic binaries with $\sim$1\,yr periods 
separated by 0\farcs8 (36\,AU) on the sky 
\citep[e.g.,][]{torres95,tokovinin99}.  Infrared measurements have shown that 
only the HD\,98800\,B system harbors a circumstellar dust disk 
\citep{koerner00,prato01}, which appears to have a nearly empty central cavity 
out to a few AU in radius \citep{furlan07}.  The coronal X-ray line ratios and 
very weak H$\alpha$ emission observed from HD\,98800\,B suggest that the stars 
are not actively accreting gas from the disk \citep{kastner04,dunkin97}.  
Hen\,3-600 appears to be a hierarchical triple star system, comprised of a 
primary spectroscopic binary of uncertain period \citep{muzerolle00,torres03} 
and a secondary visual companion located 1\farcs5 (66\,AU) to the southwest 
\citep{delareza89}.  The signatures of a circumstellar disk, including 
spectroscopic and X-ray line ratio evidence for accretion onto the central 
stars, have been identified around the Hen\,3-600\,A spectroscopic binary 
\citep[e.g.,][]{jayawardhana99,muzerolle00,huenemoerder07}.  

At their advanced ages of $\sim$10\,Myr, the nature of the circumstellar 
material in these low-mass pre-main sequence systems is a topic of active 
debate.  While the accretion signatures in the Hen\,3-600\,A 
disk are supportive of a primordial reservoir, it is unclear whether the
HD\,98800\,B disk contains gas or has evolved into a debris-only system
\citep[see the detailed discussion on this topic by][]{akeson07}.  Regardless, 
the proximity of these systems offers unusually detailed views of their 
circumstellar structures.  We aim to use the high spatial resolution afforded 
by that proximity to characterize how the disk structures have been shaped by 
their dynamical interactions with the stellar components in each system.  We 
present our observations in \S 2 and a brief qualitative analysis of the data 
in \S 3.  We discuss the radiative transfer modeling used to determine the 
disk structures in \S 4, and synthesize the results with theoretical 
predictions for disk truncation in \S 5.

\section{Observations and Data Reduction}

The HD\,98800 and Hen\,3-600 systems were observed with the Submillimeter Array 
interferometer \citep[SMA;][]{ho04} at Mauna Kea, Hawaii in the compact 
(baselines of 16-70\,m), extended (baselines of 28-226\,m), and very extended 
(baselines of 68-509\,m) configurations on 2008 March 9, February 21, and April 
3, respectively.  The SMA double-sideband receivers were tuned with a local 
oscillator (LO) frequency of 340.755\,GHz (880\,$\mu$m).  Each sideband 
provides 24 partially overlapping 104\,MHz spectral chunks centered $\pm$5\,GHz 
from the LO frequency.  The correlator was configured to subdivide one chunk 
in each sideband into 128 spectral channels to sample any emission from the 
CO $J$=3$-$2 transition (345.796\,GHz) at a velocity resolution of 0.7\,km 
s$^{-1}$.  Each of the remaining chunks was split into 32 coarser spectral 
channels to observe continuum emission.  

The observations cycled through the two targets and TW Hya (Andrews et al., 
{\it in preparation}) between integrations on the nearby quasar J1037$-$295, 
with a total cycle time of 18 minutes (5 minutes per each disk target, 3 
minutes on the quasar) in the compact and extended configurations and 12 
minutes (3 minutes for each disk target and quasar) in the very extended 
configuration.  The bright quasar 3C 279 was observed every other cycle to 
help assess the quality of phase transfer.  Additional observations of bright 
quasars (3C 84, 3C 273, 3C 279, and 3C 454.3) and Callisto were made for 
bandpass and absolute flux calibration when the targets were at low elevations 
($<$20\degr).  The observing conditions were excellent, with zenith opacities 
ranging from 0.04-0.08 (corresponding to 0.8-1.6\,mm of precipitable water 
vapor) and well-behaved phase variations on timescales longer than the 
calibration cycle.  The extended array observations (2008 February 21) were 
conducted during a configuration change, and therefore only utilized 4 of the 8 
SMA antennas.

The data were edited and calibrated with the {\tt MIR} software 
package.\footnote{See 
\url{http://cfa-www.harvard.edu/$\sim$cqi/mircook.html}.}  The 
bandpass response was calibrated with observations of bright quasars, and 
broadband continuum channels in each sideband were generated by averaging the 
central 82\,MHz in all coarse resolution spectral chunks.  The visibility 
amplitude scale was set by bootstrapping quasar flux densities from the 
observations of Callisto.  The absolute flux calibration uncertainty from this 
technique is estimated to be roughly 10\%.  The antenna-based complex gain 
response of the system as a function of time and elevation was determined with 
reference to the observations of J1037$-$295 (the nearest sufficiently bright 
calibrator), located midway between and $\sim$10\degr\ from each of the science 
targets.  The phase transfer between the calibrator and science targets over 
that angular separation is not ideal, especially given the low observing 
elevations for these southern systems.  Atmospheric decoherence on these scales 
will add phase noise that acts to smear out the emission from the targets.  
Typically, observations of a second quasar are utilized to estimate the level 
of decoherence in terms of the size of the ``seeing" disk.  Unfortunately, the 
next nearest bright quasar, 3C 279, probes the atmosphere at a much larger 
30-40\degr\ separation from the targets.  Nevertheless, the observations of 3C 
279 were used to estimate the quality of phase transfer in an extreme case and 
showed a seeing disk size of $\sim$0\farcs3.  The actual phase decoherence 
between the targets and J1037$-$295 should be much less severe, producing 
significantly smaller (but still uncertain) seeing disks.

After the gain calibration, the independent sets of visibilities from each 
configuration and sideband were compared to check for consistency on 
overlapping spatial scales.  After finding excellent agreement, all of the data 
for each target were combined.  The standard tasks of Fourier inverting the 
visibilities, deconvolution with the {\tt CLEAN} algorithm, and restoration 
with a synthesized beam were conducted with the {\tt MIRIAD} package.  Maps of 
the continuum and CO $J$=3$-$2 line emission were made by naturally weighting 
the visibilities.  The resulting continuum maps have 1\,$\sigma$ rms noise 
levels of 3.5\,mJy beam$^{-1}$, with synthesized beam dimensions of 
$0\farcs92\times0\farcs68$ at a position angle (PA, measured east of north) of 
11\degr\ for HD\,98800 and $1\farcs11\times0\farcs74$ at PA = 18\degr\ for 
Hen\,3-600.  To maximize the sensitivity to line emission, only data from the 
compact configuration were used to generate CO $J$=3$-$2 spectral images, with 
synthesized beam dimensions of $3\farcs1\times1\farcs8$ at PA = 164\degr\ and 
$4\farcs0\times1\farcs8$ at PA = 168\degr\ for HD\,98800 and Hen\,3-600, 
respectively.  No line emission was detected for either target, with 
3\,$\sigma$ upper limits of $\sim$0.6\,Jy beam$^{-1}$ in a 0.7\,km s$^{-1}$ 
channel.

\section{Results}

\begin{figure}
\epsscale{1.1}
\plottwo{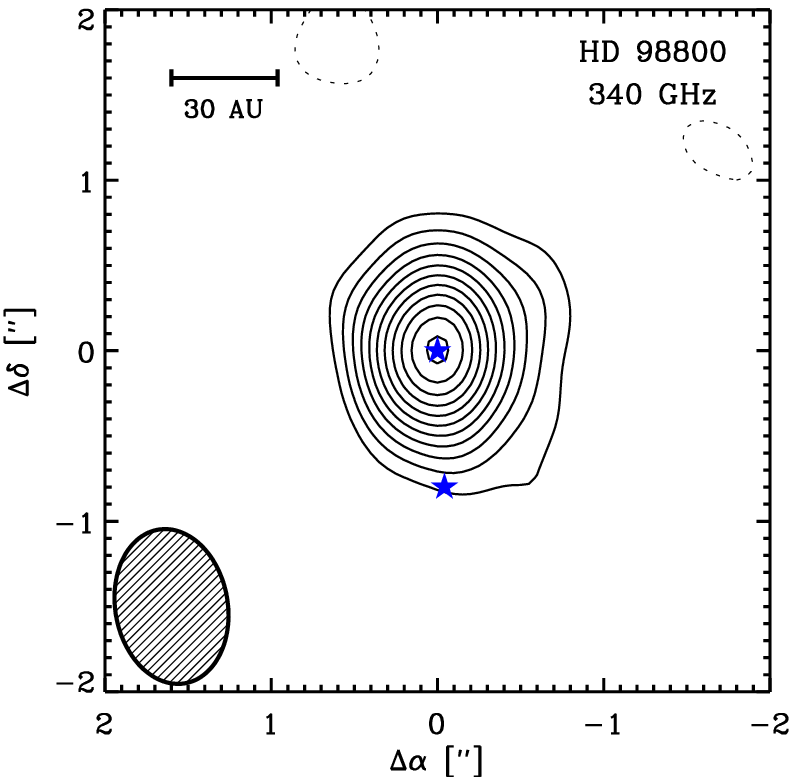}{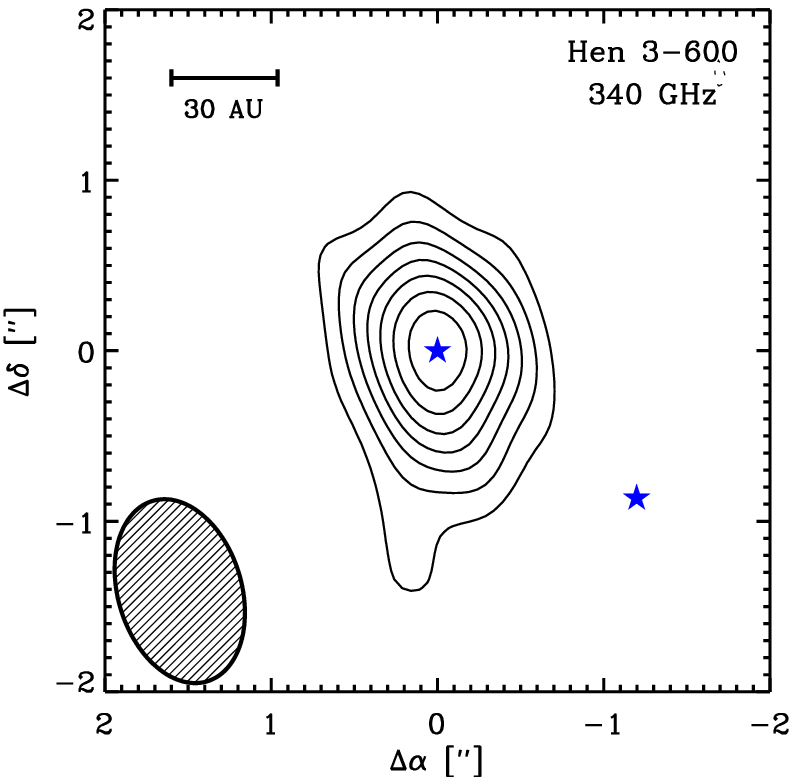}
\figcaption{Aperture synthesis images of the 880\,$\mu$m (340\,GHz) dust 
continuum emission in the HD\,98800 ({\it left}) and Hen\,3-600 ({\it right}) 
multiple star systems.  Contours start at 3\,$\sigma$ (10\,mJy beam$^{-1}$) and 
are shown at 2\,$\sigma$ intervals.  The synthesized beam dimensions are marked 
in the lower left corner of each image, and a 30\,AU scale bar is included in 
the upper left.  The images are centered at the peak of the continuum emission, 
corresponding to the positions of the HD\,98800\,B and Hen\,3-600\,A 
spectroscopic binaries.  Each stellar component is marked with a blue star 
symbol. \label{images}}
\end{figure}

The synthesized images of the 880\,$\mu$m (340\,GHz) dust continuum emission in 
the HD\,98000 and Hen\,3-600 systems are shown in Figure \ref{images}.  The 
HD\,98800 map exhibits emission with an integrated flux density of 110\,mJy and 
peak of $82\pm3$\,mJy beam$^{-1}$, centered at $\alpha = 
11^{\rm h}22^{\rm m}05\fs24$, $\delta = -24\degr46\arcmin39\farcs10$ (J2000).  
Based on the reference coordinates and proper motion estimates from {\it 
Hipparcos} \citep[which resolved the HD\,98800\,A-B system;][]{perryman97}, the 
SMA emission centroid is coincident with the HD\,98800\,B spectroscopic binary 
position to within $\sim$50\,mas, well below the SMA astrometric uncertainty of 
$\sim$0\farcs1 (and roughly on the scale of the Ba-Bb separation).  The 
location of the HD\,98800\,A spectroscopic binary is marked on Figure 
\ref{images} with a star symbol 0\farcs8 to the south of the map center.  The 
Hen\,3-600 system shows emission with an integrated flux density of 75\,mJy and 
peak of $57\pm3$\,mJy beam$^{-1}$, centered at $\alpha = 
11^{\rm h}10^{\rm m}27\fs84$, $\delta = -37\degr31\arcmin51\farcs71$ (J2000).  
Using the {\it Two Micron All Sky Survey} (2MASS) astrometry \citep{cutri03} 
and the proper motion determined by \citet{torres06}, we find that the SMA 
emission centroid is coincident with the Hen\,3-600\,A spectroscopic binary 
position to within 15\,mas in right ascension and 0\farcs15 in declination (the 
measured position is offset to the north, away from Hen\,3-600\,B).  Given the 
astrometric errors that may occur for such a low declination target as well as 
the potential confusion in the 2MASS catalog astrometry from the close 
companion, we suggest that the continuum emission measured with the SMA is 
consistent with being centered on Hen\,3-600\,A.  The location of the 
Hen\,3-600\,B companion is marked in Figure \ref{images} with a star symbol 
$\sim$1\farcs5 to the southwest of the map center.

An examination of the SMA visibilities confirms that the emission from each 
target is partially resolved (see \S 4).  An elliptical Gaussian fit to the 
HD\,98800\,B visibilities indicates an elongated emission morphology, with a 
high aspect ratio corresponding to an inclination $i \approx 55$-71\degr\ at a 
major axis orientation ${\rm PA} \approx 130$-159\degr\ east of north.  This 
inferred disk geometry is in excellent agreement with the much more precisely 
determined orbital geometry of the HD\,98800\,B spectroscopic binary, for which 
\citet{boden05} have measured $i = 67$\degr\ and $\Omega = 338$\degr\ 
(corresponding to PA = 158\degr).  Within the uncertainties of the SMA data, 
the HD\,98800\,B disk and spectroscopic binary orbital plane are aligned.  A 
similar fit to the Hen\,3-600\,A visibilities suggests a more circular emission 
morphology, with inclination estimates of $i \approx 25$-45\degr\ and a major 
axis orientation ${\rm PA} \approx 145$-185\degr\ east of north.  Lacking 
orbital information for either the spectroscopic binary or the wider visual 
pair, the relative orientation of the star and disk axes in this system remains 
unclear.

\section{Disk Models}

The 880\,$\mu$m visibilities presented here provide the first spatially 
resolved information about the HD\,98800\,B and Hen\,3-600\,A circumstellar 
disks.  We aim to use these data to directly locate the outer disk edges, where 
the circumstellar material could be truncated by tidal interactions with their 
nearby stellar companions.  Moreover, following the analysis of 
\citet{jensen97}, we seek to indirectly locate the inner disk edges sculpted by 
their central spectroscopic binaries based on the high-quality infrared spectra 
presented by \citet{uchida04} and \citet{furlan07}.  To determine these disk 
edges, we utilize a parametric truncated disk structure model and a 
two-dimensional Monte Carlo radiative transfer code \citep[{\tt RADMC}; 
see][]{dullemond04} to compute synthetic 880\,$\mu$m visibilities and spectral 
energy distributions (SEDs) that can be compared with observations.  A detailed 
description of the modeling methodology was provided by \citet{andrews09a}, and 
so here we focus only on the key modifications made to accommodate truncated 
disk structures.

We parameterize the axisymmetric density structure for the disk models based on 
two different radial zones, including a standard ``outer" disk and a central 
disk ``cavity" that contains substantially less material.  The surface 
densities have a $\Sigma \propto 1/R$ behavior for $R_{\rm cav} \le R \le 
R_{\rm out}$ in the outer zone, and a constant value (depleted by a factor 
$\delta_{\Sigma}$ from the value at $R_{\rm cav}$) inside the cavity ($R_{\rm 
in} \le R < R_{\rm cav}$).  The vertical density structure is assumed to be 
Gaussian and is parameterized by a scale-height distribution $H \propto 
R^{1+\psi}$ that can be artificially ``puffed"-up by a factor $\delta_H$ at the 
cavity edge ($R_{\rm cav}$) to account for the locally increased energy input 
from the normal irradiation of the optically thick outer disk zone 
\citep[e.g.,][]{dullemond01,dalessio05}.  The densities in these structures are 
normalized with a total dust mass, $M_{\rm dust}$.

We use the same set of dust opacities adopted by \citet{andrews09a}, computed 
for a power-law size distribution up to a maximum grain radius of 1\,mm.  For a 
small fraction, $f_{\rm sg}$, of the dust mass inside the disk cavities, the 
maximum grain size was decreased to 1\,$\mu$m (hereafter ``small" grains) to 
more faithfully reproduce the solid-state dust emission features for each 
target.  In these models, we assume the only heating source was irradiation 
from a central stellar photosphere, represented by appropriately scaled 
\citet{lejeune97} spectral templates that match the component-resolved optical 
and near infrared photometry \citep{gregorio92,soderblom98,webb99,low99,weintraub00,koerner00,prato01,geoffray01}.  Table \ref{stars_table} lists the stellar 
parameters adopted in the modeling.  The corresponding photosphere templates 
are also used to help construct the complete broadband SEDs for each disk.  
Since neither the HD\,98800\,A or Hen\,3-600\,B components are found to harbor 
any disk material, we subtract their stellar templates from the composite {\it 
Spitzer} IRS spectra for each system \citep{uchida04,furlan07} to derive the 
star+disk spectra for HD\,98800\,B and Hen\,3-600\,A alone.  Assuming that the 
emission at longer wavelengths is generated solely by the disks, we include 
the far-infrared and millimeter photometry from the literature in the SEDs 
without any such corrections \citep{jensen96,sylvester96,sylvester01,low05}.

In this modeling effort, no attempt was made to account for irradiation from 
individual components of the eccentric spectroscopic binaries or from any {\it 
external} disk irradiation by the close stellar companions 
\citep[e.g.,][]{nagel09}.  For convenience, the updated HD\,98800 {\it 
Hipparcos} distance of 45\,pc was adopted for both systems 
\citep{vanleeuwen07}.  Note that this kind of parametric modeling is still 
subject to some intrinsic degeneracies, which have been discussed at length 
elsewhere \citep{thamm94,chiang01,aw07,andrews09a}.  Some degeneracies specific
to this truncated disk structure model are highlighted below.  Because the 
880\,$\mu$m emission from the target disks is relatively faint, the exploration 
of parameter-space was limited.  Focus was placed on estimating $R_{\rm cav}$ 
and $R_{\rm out}$ for reasonable assumptions about the ranges of other input 
parameters.

\begin{figure}
\epsscale{1.1}
\plottwo{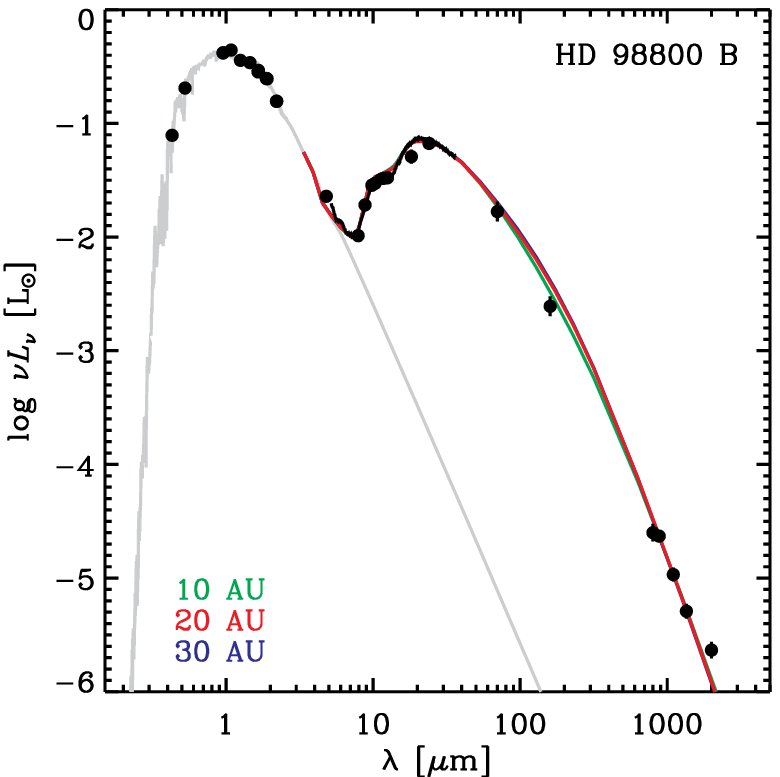}{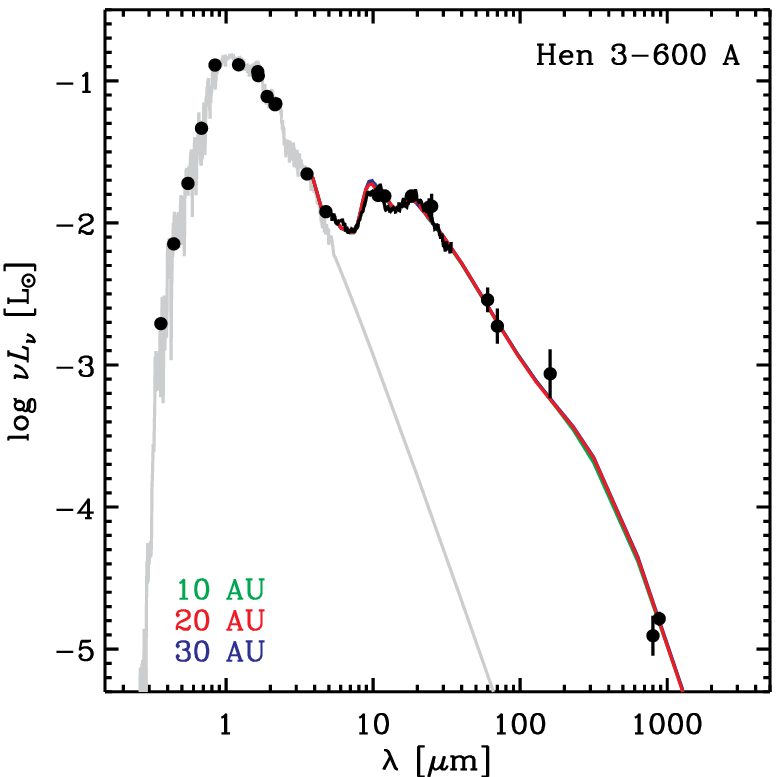}
\figcaption{Broadband spectral energy distributions for the HD\,98800 B ({\it 
left}) and Hen\,3-600\,A ({\it right}) disks, constructed from measurements in 
the literature.  Light gray curves mark the composite stellar photospheres 
adopted for the radiative transfer modeling, while the black curves show the 
{\it Spitzer} IRS spectra corrected for the contamination of the nearby 
companions (see text).  The overlaid curves are disk structure models for 
different values for the outer disk radius, $R_{\rm out}$, as labeled in the 
lower left corner of each plot.  The individual models are difficult to 
distinguish here, but show clear differences in terms of the 880\,$\mu$m 
visibilities (see Fig.~\ref{vis}).  \label{sed}}
\end{figure}

\begin{figure}
\epsscale{1.1}
\plottwo{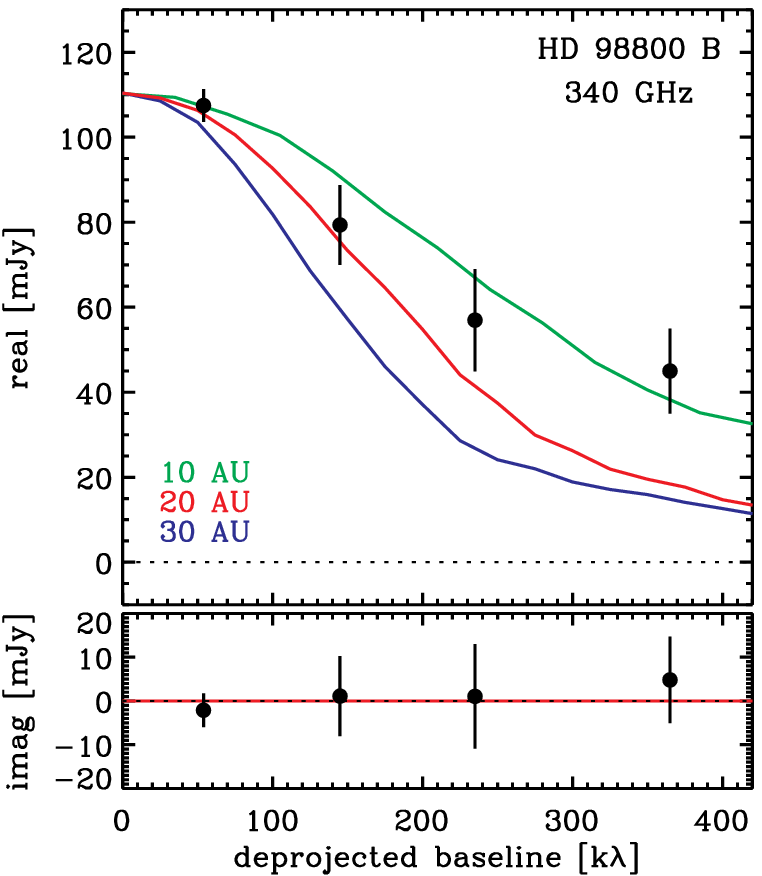}{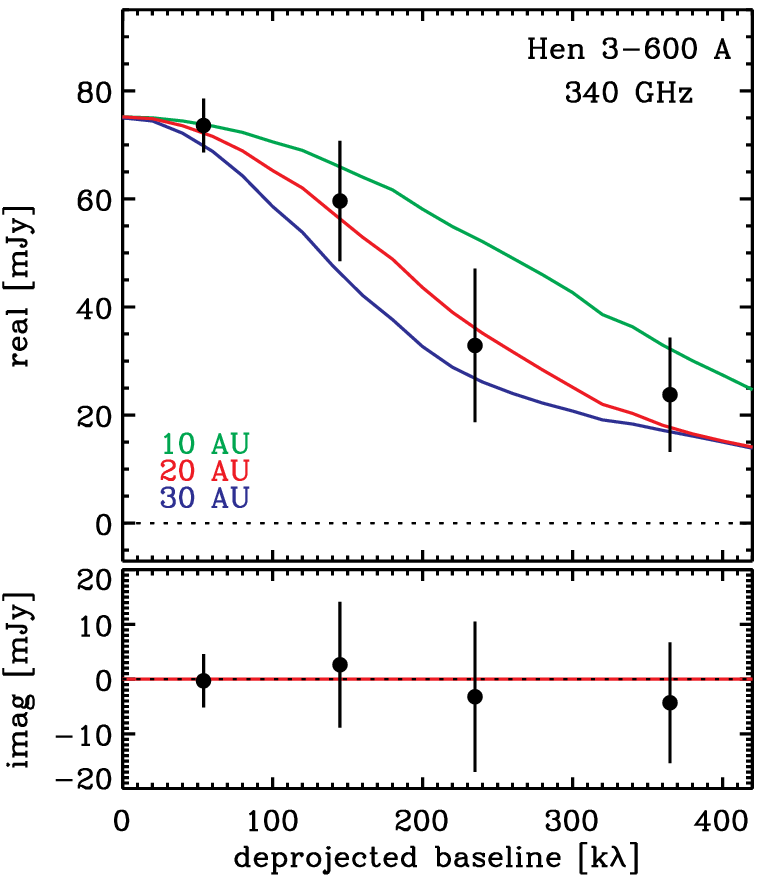}
\figcaption{The azimuthally-averaged 880\,$\mu$m continuum visibilities (real 
and imaginary parts) as a function of the deprojected baseline length for the 
HD\,98800\,B ({\it left}) and Hen\,3-600\,A ({\it right}) disks.  The overlaid 
curves are truncated disk structure models with different values for the outer 
edge radius, $R_{\rm out}$, as labeled in the lower left corner of each plot. 
\label{vis}}
\end{figure}

The best estimates for the HD\,98800\,B and Hen\,3-600\,A disk structure 
parameters described above are compiled in Table \ref{disks_table}.  The 
observed broadband SEDs and {\it Spitzer} IRS spectra from the literature are 
shown in Figure \ref{sed}, with the composite stellar photosphere spectra 
overlaid in gray.  The best-fit synthetic SEDs for disk models with several 
fixed values of the outer radius ($R_{\rm out}$) are plotted for comparison.  
Neither disk exhibits dust emission in excess of the photosphere at wavelengths 
less than $\sim$5\,$\mu$m, as was noted in previous infrared studies 
\citep[e.g.,][]{jayawardhana99,uchida04,furlan07}.  This lack of excess 
emission and the secondary emission peak near 20\,$\mu$m were reproduced in the 
models by varying the disk cavity parameters, \{$R_{\rm in}$, $R_{\rm cav}$, 
$\delta_{\Sigma}$, $\delta_H$\} (and to some extent $f_{\rm sg}$).  Of these, 
we perhaps have the best handle on $R_{\rm cav}$ due to the prominent role it 
has in shaping the rising mid-infrared spectrum and setting the wavelength of 
the secondary emission peak.  Our estimate of $R_{\rm cav} = 1$\,AU for the 
Hen\,3-600\,A disk is in good agreement with the original measurement from 
these data by \citet{uchida04}, but the $R_{\rm cav} = 3.5$\,AU inferred here 
for the HD\,98800\,B disk is somewhat smaller than the 5.9\,AU derived by 
\citet{furlan07}.  Given the differences in the adopted stellar parameters, 
dust populations, and the prescribed physical structure of the cavity edge 
between these two studies, this apparent $R_{\rm cav}$ discrepancy is probably 
irrelevant.  However, if the cavity extended out to the radius determined by 
\citet{furlan07}, we would expect to see a null in the 880\,$\mu$m visibilities 
on $\sim$350-400\,k$\lambda$ baselines, corresponding to angular scales of 
$\sim$0\farcs3 \citep[see][]{hughes07}.  Since no such null is noted in the SMA 
data (see Figure \ref{vis}), a smaller cavity size is appropriate.

The additional cavity parameters, \{$R_{\rm in}$, $\delta_{\Sigma}$, 
$\delta_H$, $f_{\rm sg}$\}, are not well-constrained and are somewhat mutually 
degenerate.  For example, one could adjust $R_{\rm in}$ and compensate for the 
modified infrared emission by varying $\delta_{\Sigma}$.  Nevertheless, in 
practice the range of viable parameter-space is relatively small, and the 
values selected here are meant only to be illustrative and facilitate 
reasonable estimates of the disk edges.  Both disks contain only a miniscule 
$\sim$10$^{-10}$\,M$_{\odot}$ of dust inside $R_{\rm cav}$, a fraction 
$f_{\rm sg}$ (by mass) of which is comprised of small grains.  The scale-height 
modifications at $R_{\rm cav}$ ($\delta_H$) are important, but also modest, 
amounting to a change from 0.19 to 0.26\,AU for HD\,98800\,B and 0.0026 to 
0.0032\,AU for Hen\,3-600\,A.  The vertical structure parameters inferred for 
both disks are similar to their non-truncated counterparts in the Ophiuchus 
star-forming region \citep{andrews09a}, but their total dust masses are an 
order of magnitude smaller than the typical T Tauri disk \citep{aw07b}.  Note 
that the Hen\,3-600\,A disk has a slightly higher $M_{\rm dust}$ than 
HD\,98800\,B, despite the latter being brighter at 880\,$\mu$m.  The apparent 
discrepancy is a result of the significantly cooler dust temperatures produced 
around the later-type Hen\,3-600\,A stars.

Because there is no spatially resolved information in the SEDs alone, those 
data can be reproduced equally well with essentially any value for the outer 
disk edge.  Not surprisingly then, the SED models for different $R_{\rm out}$ 
values shown in Figure \ref{sed} are practically identical.  However, the 
synthetic 880\,$\mu$m continuum visibilities computed from the same disk models 
exhibit distinctive signatures for different outer edge locations.  Figure 
\ref{vis} displays the azimuthally-averaged representations of the observed and 
synthetic visibilities for the HD\,98800\,B and Hen\,3-600\,A disks, 
deprojected to account for the disk viewing geometries \citep[see][]{lay97}.  
The real parts of the SMA visibilities decrease with deprojected baseline, 
demonstrating that both disks are resolved.  Moreover, the imaginary parts of 
the visibilities are essentially zero, indicating that there is no strong 
evidence for large departures from axisymmetry.  To facilitate a fair 
comparison of the data and models, the model visibilities are first convolved 
with a 0\farcs15 Gaussian to compensate for the atmospheric decoherence 
discussed in \S 2.  After this ``seeing" correction, it is clear in 
Figure \ref{vis} that the SMA data are best reproduced for $R_{\rm out}$ values
of $\sim$10-15\,AU for the HD\,98800\,B disk and $\sim$15-25\,AU for the 
Hen\,3-600\,A disk.  These values could be smaller if the decoherence is 
under-estimated, but they could only be larger if $\Sigma$ were a steeper 
function of radius \citep[see][]{aw07}.

\section{Discussion}

Through radiative transfer modeling of resolved 880\,$\mu$m continuum emission 
and pan-chromatic spectral energy distributions, we have characterized the key 
properties of the circumstellar material in two of the nearest pre-main 
sequence multiple star systems.  In principle, the disk structures determined 
here can be used to help assess the potential for planet formation in the 
presence of a stellar companion and to test theoretical calculations of 
star-disk tidal interactions.  The latter task has further implications for 
disk studies, as empirical constraints on these dynamical interactions can be 
used as high mass ratio ($\mu$) touchstones for modeling the impacts of young 
giant planet companions on their local disk environments.

Unfortunately, very little is known about the stellar orbits in the Hen\,3-600 
hierarchical triple system.  \citet{muzerolle00} noted that the primary appears 
to be a double-lined spectroscopic binary with a large radial velocity 
difference ($\sim$40\,km s$^{-1}$), although repeated measurements have not 
been able to extract a reliable set of orbital parameters \citep{torres03}.  If 
the inner truncation radius estimated for the Hen\,3-600\,A disk, $R_{\rm cav} 
\approx 1$\,AU, is indeed set by the dynamical clearing effects of the central 
spectroscopic binary, the calculations by \citet{artymowicz94} predict that the 
orbital separation should be in the range $a \approx 0.2$-0.6\,AU, depending on 
the eccentricity (mass ratio-related effects are small when $\mu \approx 
0.1$-0.5).  The visual companion Hen\,3-600\,B has been located at angular 
separations of 1.4-1\farcs5 to the southwest (${\rm PA} \approx 215$-230\degr) 
of the spectroscopic binary system, corresponding to projected spatial 
separations of 63-67\,AU \citep{delareza89,reipurth93,webb99,brandeker03}.  
While there are no available constraints on the A-B orbit, projection effects 
should only produce up to a $\sim$20\%\ uncertainty in the physical separation 
if the disk plane and orbit are aligned.  Assuming that the similar spectral 
types of the A and B components imply a roughly equal-mass system ($\mu \approx 
0.5$), tidal interactions would externally truncate the Hen\,3-600\,A disk at a 
radius of $\sim$25\,AU for a circular orbit \citep{artymowicz94}.  The outer 
disk edge inferred here, $R_{\rm out} \approx 15$-25\,AU, is in good agreement 
with that prediction, with the uncertainty permitting moderate orbital 
eccentricities.  

\begin{figure}
\epsscale{0.5}
\plotone{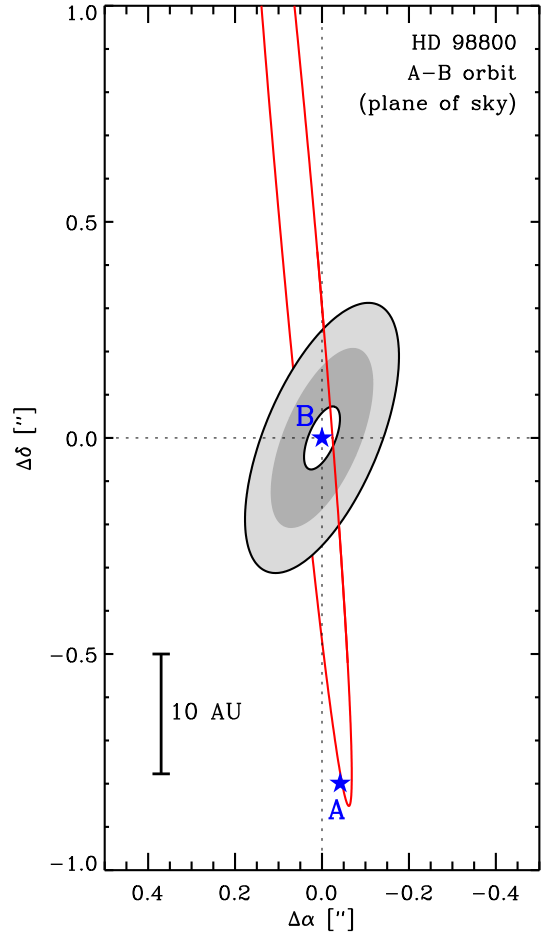}
\figcaption{The inferred geometry of the HD\,98800\,B disk ({\it grayscale}) is 
compared with the A-B visual binary orbit ({\it red}) computed by 
\citet{tokovinin99}, projected onto the plane of the sky.  The two-toned gray 
shading of the disk corresponds to the quoted range of possible $R_{\rm out}$ 
values ($\sim$10-15\,AU).  \label{orbit}}
\end{figure}

The stellar orbital parameters in the HD\,98800 quadruple system are 
comparatively more robust, thanks to long-term and intensive monitoring 
efforts.  Most recently, \citet{boden05} combined infrared interferometric 
observations with the extant spectroscopic data from \citet{torres95} to 
measure the orbital elements and individual stellar properties in the 
HD\,98800\,B spectroscopic binary.  According to the simulations of 
\citet{pichardo08}, this nearly equal-mass ($\mu = 0.43$), eccentric ($e = 
0.78$), and close ($a = 1.05$\,AU) system can harbor a circumbinary disk with 
an innermost stable orbit at a radius of $\sim$3.7\,AU.  A similar value can be 
deduced from the calculations of \citet{holman99}, or from a slight 
extrapolation of the results of \citet{artymowicz94}.  That value is 
practically identical to the internal truncation radius determined from the 
infrared spectrum morphology, $R_{\rm cav} = 3.5$\,AU.  For the wider HD\,98800 
A-B visual pair, \citet{torres95} compiled a relative astrometric record of the 
resolved system over the past century, along with a single-epoch measurement of 
the systemic radial velocity difference between the components.  With that 
information and the {\it Hipparcos} parallax, \citet{tokovinin99} determined a 
family of acceptable A-B orbits with $\sim$300-430\,yr periods, moderate 
eccentricities ($e = 0.3$-0.6), and $a \approx 50$-80\,AU.  For reference, the 
relative geometries of the HD\,98800\,B disk and an example A-B orbit from 
\citet{tokovinin99} are shown together in Figure \ref{orbit}, projected onto 
the plane of the sky.  Note that the A-B orbital plane ($i 
\approx 88$\degr, $\Omega \approx 185$\degr) is not aligned with the B 
spectroscopic binary and disk plane ($i \approx 67$\degr, $\Omega \approx 
338\degr$), making a quantitative prediction for the external disk truncation 
difficult.  Although the aforementioned theoretical calculations all 
assume coplanar interactions, we can use them as a guide to assess whether or 
not the outer truncation radius determined here is reasonable.  Assuming that 
the A and B spectroscopic binaries are roughly equal in total mass, the models 
of \citet{artymowicz94} or \citet{pichardo05} predict that the HD\,98800\,B 
disk should be truncated at a radius of $\sim$8-20\,AU, depending on the 
assumed disk viscosity and which \citet{tokovinin99} orbit is adopted.  Those 
values are indeed consistent with the estimates based on modeling the resolved 
SMA data, where $R_{\rm out} \approx 10$-15\,AU.

In general, the locations of the disk edges derived from our radiative transfer 
modeling are in good agreement with the theoretical predictions of disk 
truncation due to star-disk tidal interactions.  However, there are some 
notable problems in the details that should be brought to attention.  One of 
these involves the proper interpretation of the inner disk radius, $R_{\rm 
in}$, and the small amount of dust between it and the cavity edge 
($R_{\rm cav}$).  For the disk models used here, that small dust mass is 
necessary to reproduce the weak infrared excesses and silicate emission 
features present in the observed SEDs.  However, the dynamical models suggest 
that there are no long-term stable dust orbits inside the inferred 
$R_{\rm cav}$.  Perhaps the simplest way of reconciling this discrepancy is to 
argue that the adopted parameterization of the inner disk structure is 
inappropriate or incomplete.  For example, similar infrared excess features 
could potentially be reproduced by modifying the dust content, shape, or 
symmetry of the disk at the cavity edge \citep[e.g.,][]{isella05,akeson07}.  
With the same data for the HD\,98800\,B disk, \citet{furlan07} employed a 
narrow ring interior to their slightly larger cavity edge to reproduce the 
infrared spectrum with equivalent accuracy.  The ability to obtain quality fits 
to the observations with such different interior structures highlights the 
model degeneracies, which will remain until the inner disk edge is spatially 
resolved. 

But if the inner disk structures determined here are taken at face value, the 
material inferred to be inside the cavity edge could be interpreted as 
dynamically unstable, perhaps representative of a low-density inward flow due 
to Poynting-Robertson drag or disk accretion \citep[e.g., see][]{pierens08}.  
\citet{furlan07} suggested the former mechanism to explain the narrow ring from 
1.5-2\,AU they inferred in the HD\,98800\,B system, but rightly acknowledged 
that the empty region from 2\,AU to their $R_{\rm cav}$ (5.9\,AU) was 
incompatible with such a flow (note also that a significant fraction of such a 
ring would be {\it inside} the spectroscopic binary orbit).  The structure 
parameterization adopted here is continuous from $R_{\rm cav}$ inwards to 
$R_{\rm in}$ (which corresponds to the spectroscopic binary apastron distance), 
and so avoids this complication.  Therefore, P-R drag could potentially account 
for the material inside the stable orbit radius in this disk if there is a 
continually replenished dust supply near $R_{\rm cav}$ and viscous gas drag 
forces are negligible.  At least the latter condition appears to be met for the 
HD\,98800\,B disk, where no evidence for accretion or molecular gas has been 
uncovered \citep{dent05,salyk09}.  The same is not true for the Hen\,3-600\,A 
disk, which exhibits the clear signatures of a $\dot{M} \approx 
5\times10^{-11}$\,M$_{\odot}$ yr$^{-1}$ gaseous accretion flow from the outer 
disk to the spectroscopic binary \citep{muzerolle00}.  In this case, the small 
amount of material inside $R_{\rm cav}$ may be tracing dust grains swept up in 
that flow.  

When future astrometric and spectroscopic monitoring has fully determined the 
visual orbits in the HD\,98800 and Hen\,3-600 systems, the lack of millimeter 
emission around the secondaries can also help constrain how their circumstellar 
material evolved.  In principle, the tidal truncation calculations described 
above predict similar (perhaps slightly smaller) outer edge locations for the 
disks around the nearly equal-mass secondaries 
\citep{artymowicz94,pichardo05}.  If the individual disks initially had the 
same masses, the similar disk sizes would have produced comparable millimeter 
emission levels from both components.  Obviously this is not the case for these 
targets, nor for a handful of other resolved multiple systems 
\citep{jensen03,patience08}.  This could be an artifact of the multiple star 
formation process itself, as \citet{bate97} have argued that competitive 
accretion will produce higher initial disk masses around primary stars 
\citep[see also][]{bate00}.  Alternatively, the dynamical interactions in these 
systems could be considerably more complicated than the standard tidal 
truncation calculations have assumed, with the secondary disks suffering more 
severe disruption.  For example, the interactions in non-coplanar systems like 
the HD\,98800\,A-B visual binary could be catastrophic, producing a much more 
complex circumstellar structure \citep{verrier08}.  

The Hen\,3-600 and HD\,98800 multiple star systems should continue to be 
exploited as essential test cases for evaluating how external dynamical 
interactions can contribute to disk evolution, and therefore affect the 
potential for planet formation in the presence of stellar companions.  The 
unique proximity of these particular systems is crucial, providing both high 
sensitivity and spatial resolution to circumstellar dust structures as well as 
rare opportunities to determine precise orbits for stellar pairs with different 
separations within the same system.  While this study has tackled the initial 
challenge of resolving the millimeter emission in these systems and relating it 
to the global disk structures, the real utility of these targets now lies with 
continued work that focuses on the details.  For example, future observations 
of these disks with the {\it Atacama Large Millimeter Array} could be sensitive 
to the structural asymmetries or recently ejected material streams expected 
from non-coplanar tidal interactions \citep{akeson07,verrier08}.  However, it 
is equally important to calculate the stellar orbits in these nearby systems, 
particularly in the less-studied case of Hen\,3-600.  In fact, there are 
exciting opportunities to simultaneously refine the stellar orbits and directly 
probe star-disk interactions for the HD\,98800 system, as the visual A-B binary 
will reach periastron in the next $\sim$15\,years \citep{tokovinin99}.  
Therefore, with continued monitoring the circumstellar material in the 
Hen\,3-600 and HD\,98800 systems can eventually serve as the primary 
observational calibrators for models of star-disk (as well as planet-disk) 
tidal interactions.

\acknowledgments We are grateful to David Latham and Guillermo Torres for 
sharing their orbital expertise on these systems, and especially to Elise 
Furlan and the {\it Spitzer} GTO team lead by Dan Watson for kindly permitting 
the use of their calibrated IRS spectra.  We thank the referee for an 
insightful review that helped improve the article.  The SMA is a joint project 
between the Smithsonian Astrophysical Observatory and the Academia Sinica 
Institute of Astronomy and Astrophysics and is funded by the Smithsonian 
Institute and the Academia Sinica.  Support for this work was provided by NASA 
through Hubble Fellowship grant HF-01203.01-A awarded by the Space Telescope 
Science Institute, which is operated by the Association of Universities for 
Research in Astronomy, Inc., for NASA, under contract NAS 5-26555.  Partial 
support for I.~C.~was provided by the NSF REU and DOD ASSURE programs under NSF 
grant 0754568 and by the Smithsonian Institution.  D.~J.~W.~acknowledges 
support from NASA Origins grant NNG05GI81G.  C.~E.~was supported by the 
National Science Foundation under award 0901947.

\begin{deluxetable}{lcccc}
\tablecolumns{5}
\tablewidth{0pc}
\tablecaption{Adopted Stellar Parameters\label{stars_table}}
\tablehead{
\colhead{Star} & \colhead{$T_{\rm eff}$} & \colhead{$R_{\ast}$} & \colhead{$\log{g}$} & \colhead{$A_V$} \\
\colhead{} & \colhead{[K]} & \colhead{[R$_{\odot}$]} & \colhead{[cm s$^{-2}$]} & \colhead{[mags]}}
\startdata
HD\,98800\,Aa+Ab  & 4500 & 1.30 & 4.00 & 0.0 \\
HD\,98800\,Ba     & 4250 & 1.16 & 4.25 & 0.3 \\
HD\,98800\,Bb     & 4000 & 0.90 & 4.25 & 0.3 \\
\hline
Hen\,3-600\,Aa+Ab & 3350 & 1.20 & 4.50 & 0.0 \\
Hen\,3-600\,B     & 3350 & 1.00 & 4.50 & 0.0 
\enddata
\tablecomments{The effective temperatures, radii, gravities, and extinctions 
determined for each stellar component, based on matching resolved 
optical/near-infrared photometry to scaled \citet{lejeune97} spectral 
templates.  The parameters for HD\,98800\,B are based on the \citet{boden05} 
measurements: the templates were summed into a 
composite spectrum for the radiative transfer modeling.  The $A_V$ estimates 
correspond to the \citet{mathis90} extinction law.}
\end{deluxetable}

\begin{deluxetable}{lcc}
\tablecolumns{3}
\tablewidth{0pc}
\tablecaption{Estimated Disk Parameters\label{disks_table}}
\tablehead{
\colhead{Parameter} & \colhead{HD\,98800\,B} & \colhead{Hen\,3-600\,A}}
\startdata
$R_{\rm in}$ [AU]     & 2.0              & 0.4              \\
$R_{\rm cav}$ [AU]    & 3.5              & 1.0              \\
$R_{\rm out}$ [AU]    & 10-15            & 15-25            \\
$M_{\rm dust}$ [M$_{\odot}$]   & $3\times10^{-6}$ & $7\times10^{-6}$ \\
$\delta_\Sigma$       & $\sim$1000       & $\sim$500        \\
$H_{\rm 10\,AU}$ [AU] & 0.5              & 0.3              \\
$\psi$                & 0.10             & 0.07             \\
$\delta_H$            & 1.4              & 1.2              \\
$f_{\rm sg}$          & 0.25             & 0.10             \\
$i$ [\degr]           & 67               & 36               \\
PA [\degr]            & 158              & 169              \\
\enddata
\tablecomments{Disk structure parameters estimated from the radiative transfer 
modeling.  The parameter definitions are given in \S 4.}
\end{deluxetable}

\end{document}